\documentclass[aps, twocolumn, letterpaper, superscriptaddress]{revtex4}

\usepackage{amsmath}
\usepackage{amssymb}
\usepackage{mathrsfs}
\usepackage[dvipdfmx]{graphicx}
\usepackage{color}
\usepackage{xspace}
\usepackage{bbold}

\newcommand{\ie}{{i.e.,\/}\xspace}

\newcommand{\shout}[1]{\textcolor{black}{#1}}

\newcommand{\eq}[1]{(\ref{#1})}
\newcommand{\Eq}[1]{Eq.~(\ref{#1})}
\newcommand{\Eqs}[1]{Eqs.~(\ref{#1})}
\newcommand{\Fig}[1]{Fig.~\ref{#1}}
\newcommand{\Figs}[1]{Figs.~\ref{#1}}
\newcommand{\Sec}[1]{Sec.~\ref{#1}}
\newcommand{\Ref}[1]{Ref.~\cite{#1}}
\newcommand{\Refs}[1]{Refs.~\cite{#1}}

\newcommand{\mc}[1]{\mathcal{#1}}
\newcommand{\mcc}[1]{\mathfrak{#1}}
\newcommand{\msf}[1]{\mathsf{#1}}

\newcommand{\pd}{\partial}
\newcommand{\del}{\nabla}
\renewcommand{\vec}[1]{\boldsymbol{\rm #1}}
\newcommand{\oper}[1]{\smash{\widehat{#1}}}
\newcommand{\dd}{\mathrm{d}}

\newcommand{\Deff}{\msf{D}}
\newcommand{\sigmap}{{\bar{\sigma}}}
\newcommand{\dk}{\pi}

\newcommand{\parade}{\textit{\mbox{PARADE}}\xspace}

\begin{document}

\title{Quasioptical modeling of wave beams with and without mode conversion:\\II. Numerical simulations of single-mode beams}

\author{K. Yanagihara}
\affiliation{Nagoya University, 464-8601, Nagoya, Aichi, Japan}

\author{I. Y. Dodin}
\affiliation{Princeton Plasma Physics Laboratory, Princeton, New Jersey, 08543, USA}

\author{S. Kubo}
\affiliation{Nagoya University, 464-8601, Nagoya, Aichi, Japan}
\affiliation{National Institute for Fusion Science, National Institutes of Natural Sciences, 509-5292, Toki, Gifu, Japan}

\date{\today}

\begin{abstract}
This work continues a series of papers where we propose an algorithm for quasioptical modeling of electromagnetic beams with and without mode conversion. The general theory was reported in the first paper of this series, where a parabolic partial differential equation was derived for the field envelope that may contain one or multiple modes with close group velocities. Here, we present a corresponding code \parade (PAraxial RAy DEscription) and its test applications to single-mode beams in vacuum and also in inhomogeneous magnetized plasma. The numerical results are compared, respectively, with analytic formulas from Gaussian-beam optics and also with cold-plasma ray tracing. Quasioptical simulations of mode-converting beams are reported in the next, third paper of this series.
\end{abstract}

\maketitle
\bibliographystyle{full}

\section{Introduction}
\label{sec:intro}

Electron-cyclotron heating and current drive in fusion plasmas involve electron-cyclotron waves (ECWs) and require deposition of the wave power with high precision; hence, accurate modeling of ECWs is required. Full-wave simulations of ECWs \cite{foot:vdovin06} are computationally-demanding and largely impractical due to the fact that the ECW wavelengths (which are in the mm range) are extremely small compared to the plasma size. Geometrical-optics (GO) ray-tracing simulations \cite{ref:tsujimura15, ref:marushchenko14} are much faster but cannot adequately describe the structure of the wave beam in the focal region. Many alternatives have been proposed to improve reduced modeling of ECWs, including complex-eikonal methods \cite{ref:mazzucato89, ref:nowak93, ref:peeters96, ref:farina07}, beam tracing \cite{ref:pereverzev98, ref:poli01b, ref:poli01, ref:poli18}, and the quasioptical approach \cite{ref:balakin08b, ref:balakin07a, ref:balakin07b}. However, none of them account for the possible interaction of the two cold-plasma modes (linear mode conversion), which can occur in strongly sheared magnetic field at low density, for example, at the plasma edge \cite{my:xo, ref:kubo15}. Hence, a need remains for a more systematic approach to modeling ECWs.

This work continues a series of papers where we propose a new algorithm for quasioptical modeling of electromagnetic (EM) beams propagating in inhomogeneous plasma with and without mode conversion. \shout{Although our project is largely motivated by the need to improve ECW modeling in fusion plasmas, our results are equally applicable also beyond plasma physics, for example, to studying related problems in optics and general relativity \cite{ref:bliokh15, foot:marius}.} The general theory was reported in Paper~I of our series \cite{foot:paper1}. There, a parabolic partial differential equation was derived for a certain projection of the field envelope, whose dimension $N \ge 1$ equals the number of the resonantly-coupled modes. Based on that theory, which is a generalization of extended geometrical optics (XGO) \cite{phd:ruiz17, my:covar, my:qdirac, my:qdiel}, we have also developed a code \parade (PAraxial RAy DEscription) and applied it to several test problems. Paper~III of our series \cite{foot:paper3} reports \parade's applications to quasioptical simulations of mode-converting beams with $N = 2$. Here, we present a simplified version of the same code for $N = 1$, in which case mode conversion is not considered. This ``scalar'' version of \parade serves two goals: (i) it is a stepping stone towards a more general code reported in Paper~III; and (ii) it is useful as an independent tool for quasioptical beam tracing without mode conversion in sufficiently dense plasma, especially when the absorption is strongly inhomogeneous and no simple ansatz can be assumed for the beam structure. \shout{(}In this second capacity, \parade can also be considered as an alternative to the algorithm proposed in \Refs{ref:balakin07a, ref:balakin07b, ref:balakin08b}. The latter is similar in spirit \cite{foot:balakincomp} but is limited to single-mode beams by design.\shout{) But fusion applications of \parade \cite{foot:fusion} will not be discussed in this paper, which is focused on the general method rather than on physics of specific waves. Instead, we shall present examples of \parade simulations in the basic-physics context, namely, by modeling beam propagation in vacuum and in cold magnetized plasma. Simple geometries will be assumed in order to benchmark the code and illustrate its features most transparently.}

This paper is organized as follows. In \Sec{sec:xgo}, we introduce the key equations derived in Paper~I and also adjust them to numerical modeling. In \Sec{sec:code}, we outline the numerical algorithm used in \parade. In \Sec{sec:sim}, we report simulation results for test problems. In \Sec{sec:conc}, we summarize our main conclusions.

\section{Theoretical model}
\label{sec:xgo}

\subsection{Basic equations}
\label{sec:basic}

Here, we outline how the general theory developed in Paper~I can be applied, with some adjustments, to describing single-mode \shout{wave beams}. We start by assuming a general linear equation for the electric field~$\vec{E}$ of a wave,
\begin{gather}\label{eq:E}
\oper{\vec{D}} \vec{E} = 0,
\end{gather}
where $\oper{\vec{D}}$ is a linear dispersion operator discussed below. We also assume that the field can be represented in the eikonal form,
\begin{gather}\label{eq:psi}
\vec{E} = \vec{\psi} e^{i\theta},
\end{gather}
where $\vec{\psi}$ is a slow complex vector envelope and $\theta$ is a fast real ``reference phase'' to be prescribed (\Sec{sec:basic}). The wave is considered stationary, so it has a constant frequency $\omega$; then $\vec{\psi}$ and $\theta$ are functions of the spatial coordinate $\vec{x}$. Correspondingly, the envelope $\vec{\psi}$ satisfies
\begin{gather}
\oper{\vec{\mc{D}}} \vec{\psi} = 0,
\quad
\oper{\vec{\mc{D}}} \doteq e^{-i\theta} \oper{\vec{D}} e^{i\theta},
\end{gather}
where $\oper{\vec{\mc{D}}}$ serves as the ``envelope dispersion operator'' ($\doteq$ denotes definitions).
We introduce $\vec{k} \doteq \del \theta$ for the local wave vector, $\lambda \doteq 2\pi/k$ for the wavelength, $L_\parallel$ for the characteristic scale of the beam field $\vec{\psi}$ along the group velocity at the beam center, and $L_\perp$ for the minimum scale of the field in the plane transverse to this group velocity. The medium-inhomogeneity scale is assumed to be larger than or comparable with $L_\parallel$.

We adopt the quasioptical ordering, specifically,
\begin{gather}\label{eq:eps}
\epsilon_\parallel \doteq \lambda/L_\parallel,
\quad
\epsilon_\perp \doteq \lambda/L_\perp,
\quad
\epsilon_\parallel \sim \epsilon_\perp^2 \ll 1.
\end{gather}
Using this ordering, the ``envelope dispersion operator'' $\oper{\vec{\mc{D}}}$ can be expressed as
\begin{gather}\label{eq:envD}
\oper{\vec{\mc{D}}} \vec{\psi} = \vec{\Deff} \vec{\psi} + \oper{\vec{\mc{L}}}_\epsilon \vec{\psi}.
\end{gather}
The operator $\oper{\vec{\mc{L}}}_\epsilon = O(\epsilon_\perp)$ is specified in Paper~I (also see below). The matrix $\vec{\Deff}$ serves as the ``effective dispersion tensor'' found from $\oper{\vec{D}}$, and satisfies the ordering
\begin{gather}\label{eq:DHDA}
\vec{\Deff}_H = O(1), \quad \vec{\Deff}_A = O(\epsilon_\parallel),
\end{gather}
where the indices $H$ and $A$ denote the Hermitian and anti-Hermitian parts, correspondingly. Assuming that spatial dispersion is weak enough, $\vec{\Deff}$ can be replaced \cite{foot:paper1} with the homogeneous-plasma dispersion tensor,
\begin{gather}\label{eq:D}
\vec{D}(\vec{x}, \vec{p})
 = \frac{c^2}{16\pi \omega^2}\,[\vec{p}\vec{p} - (\vec{p} \cdot \vec{p})\mathbb{1}]
 + \frac{1}{16\pi}\,\vec{\varepsilon}(\vec{x}, \vec{p}),
\end{gather}
where $\mathbb{1}$ is a unit matrix and $\vec{\varepsilon}$ is the homogeneous-plasma dielectric tensor \cite{book:stix}; its dependence on $\omega$ is assumed but not emphasized, since $\omega$ is constant. (Here, $\vec{p}$ denotes any given wave vector, as opposed to $\vec{k}$, which is the specific wave vector determined by $\theta$; see above.) Note that \Eq{eq:D} assumes the Euclidean metric. Although we shall also use curvilinear coordinates below, expressing $\vec{D}$ in those coordinates will not be necessary as explained in Sec.~VI\,D of Paper~I.

With $\vec{\Deff} \approx \vec{D}$ and $\vec{D}$ given by \Eq{eq:D}, \Eq{eq:DHDA} implies $\vec{\varepsilon}_A = O(\epsilon_\parallel)$. Hence, $\vec{D}_H$ is the dominant part of $\vec{D}$, and \shout{\Eq{eq:envD}} yields
\begin{gather}\label{eq:Dpsi}
\vec{D}_H \vec{\psi} = O(\epsilon_\perp).
\end{gather}
As a Hermitian matrix, $\vec{D}_H$ has enough (three) eigenvectors $\vec{\eta}_s$ to form a complete orthonormal basis. Then, it is convenient to represent the envelope in this basis,
\begin{gather}
\vec{\psi} = \vec{\eta}_s a^s,
\end{gather}
where $a^s$ are the corresponding complex amplitudes. [Summation over \shout{repeated} indices is assumed. For all functions derived from $\vec{D}$, such as $\vec{\eta}_s$ here, the notation convention $f \equiv f(\vec{x}) \equiv f(\vec{x}, \vec{k}(\vec{x}))$ will also be assumed by default.] Then,
\begin{gather}
\vec{D}_H \vec{\psi} =  \vec{\eta}_s \Lambda_s a^s,
\quad
\vec{D}_H \vec{\eta}_s =  \Lambda_s \vec{\eta}_s,
\end{gather}
where $\Lambda_s$ are the corresponding eigenvalues. Due to \Eq{eq:Dpsi} and the mutual orthogonality of all $\vec{\eta}_s$, this means that $\Lambda_s a^s = O(\epsilon_\perp)$ for every given $s$ individually; hence, either $a^s$ is small or $\Lambda_s$ is small. In the latter case, $\vec{\eta}_s$ approximately satisfies the eigenmode equation, ${\vec{D}_H \vec{\eta}_s = \Lambda_s \vec{\eta}_s \approx 0}$, so it can be viewed as the local polarization vector of a GO mode. Then, the corresponding $a^s$ can be understood as the local scalar amplitude of an actual GO mode, so that $a^s = O(1)$ is allowed.
In this paper, we assume conditions such that \textit{only one mode} can be excited at given $\vec{x}$ near the fixed frequency of interest. Then,
\begin{gather}\label{eq:psia}
\vec{\psi} = \vec{\Xi} a + O(\epsilon_\perp),
\quad
\vec{D}_H \vec{\Xi} = \vec{\Xi} \Lambda,
\end{gather}
where the polarization vector $\vec{\Xi}$ is the eigenvector $\vec{\eta}_s$ corresponding to the excited mode and $\Lambda$ is the corresponding eigenvalue. (The notation is chosen here such that our formulas extend to mode-converting waves with generalized definitions of $\vec{\Xi}$, $\Lambda$, and $a$ as in \Refs{foot:paper1, foot:paper3}.)
The $O(\epsilon_\perp)$ term in \Eq{eq:psia} can be easily calculated from $a$ and is included in our theory as a perturbation \cite{foot:paper1} but will not be considered below explicitly.
Hence, we shall describe the wave in terms \shout{of the scalar $a$}, which can also be expressed as follows:
\begin{gather}
a = \vec{\Xi}^+ \vec{\psi}.
\end{gather}
Here, the row vector $\vec{\Xi}^+$ is the dual polarization vector, so $\vec{\Xi}^+ \vec{\Xi} = 1$. Since we consider the beam dynamics in coordinates that are close to Euclidean, it will be sufficient, within the accuracy of our model, to consider the row vector $\vec{\Xi}^+$ simply as the conjugate transpose of the column vector $\vec{\Xi}$.

\subsection{Reference ray}

We shall consider the beam evolution in the coordinate system that is determined by a family of rays. These rays are governed by Hamilton's equations \cite{foot:paper1}
\begin{gather}\label{eq:rtp}
\frac{\dd x^\alpha}{\dd \tau} = \frac{\pd\Lambda(\vec{x}, \vec{p})}{\pd p_\alpha},
\quad
\frac{\dd p_\alpha}{\dd\tau} = - \frac{\pd \Lambda(\vec{x}, \vec{p})}{\pd x^\alpha},
\end{gather}
where $\tau$ serves as an effective time variable, $x^\alpha$ denote the ray location in the laboratory coordinates, and $p_\alpha$ denote the ray wave-vector components in the associated momentum space. The initial conditions are assumed such that $\Lambda(\vec{x}(0), \vec{p}(0))$ is of order $\epsilon_\parallel$ or smaller \cite{foot:paper1}. (A more specific definition will not be needed.) Then, it also remains small indefinitely, because $\Lambda$ is conserved by \Eqs{eq:rtp}.

Let us introduce the path along each ray, $\zeta \doteq \int_0^\tau V\,d\tau$, where $V \doteq |\vec{V}|$ and $\vec{V}$ is given by
\begin{gather}
V^\alpha(\vec{x}, \vec{p}) \doteq \frac{\pd\Lambda(\vec{x}, \vec{p})}{\pd p_\alpha}.
\end{gather}
At any given $\zeta$, the coordinates of rays with different initial $\vec{x}$ form a ``transverse space''. Let us define a two-dimensional transverse coordinate $\tilde{\vec{\varrho}} \equiv \{\tilde{\varrho}^1, \tilde{\varrho}^2\}$ on this space such that its origin corresponds to the location of some ``reference ray'' (RR). This ray can be chosen arbitrarily as long as it is representative of the rays within the beam. The coordinate and the wave vector of the RR are denoted as $\vec{X}$ and $\vec{K}$, respectively. Like the coordinate and momentum of other rays [\Eqs{eq:rtp}], they can be found by solving the ray equations
\begin{gather}\label{eq:rtpX}
\frac{\dd X^\alpha}{\dd\zeta} = \frac{V^{\alpha}_\star}{V_\star},
\quad
\frac{\dd K_\alpha}{\dd\zeta} = - \frac{1}{V_\star}\,\frac{\pd \Lambda_\star}{\pd X^\alpha},
\end{gather}
where the index $\star$ denotes evaluation at $(\vec{X}, \vec{K})$. Note that we require $\Lambda_\star$ to be exactly zero initially; then, it also remains zero at all $\zeta$,
\begin{gather}\label{eq:Lz}
\Lambda_\star = 0.
\end{gather}

\subsection{Ray-based coordinates}

Since there is a one-to-one correspondence between the laboratory coordinates $x^\alpha \equiv \{x^1, x^2, x^3\}$ and $\tilde{x}^\mu \equiv \{\zeta, \tilde{\varrho}^1, \tilde{\varrho}^2\}$, one can choose the latter as the new coordinates. The two sets of coordinates are connected via
\begin{gather}
\dd \vec{x} = \tilde{\vec{e}}_\mu \dd\tilde{x}^\mu,
\end{gather}
where $\tilde{\vec{e}}_\mu$ are the basis vectors of the ``tilded'' coordinate system. Accordingly, the transformation matrices $\vec{\mc{X}}$ and
\begin{gather}\label{eq:tX}
 \tilde{\vec{\mc{X}}} \doteq \vec{\mc{X}}^{-1},
\end{gather}
where $\mc{X}^\alpha{}_\mu \doteq (\tilde{\vec{e}}_\mu)^\alpha$ are given by
\begin{gather}\label{eq:XX}
\mc{X}^\alpha{}_\mu = \frac{\pd x^\alpha}{\pd \tilde{x}^\mu},
\quad
\tilde{\mc{X}}^\mu{}_\alpha = \frac{\pd \tilde{x}^\mu}{\pd x^\alpha}.
\end{gather}

We choose the transverse coordinates such that
\begin{gather}\label{eq:eee}
\tilde{\vec{e}}_{\star\mu} \cdot \tilde{\vec{e}}_{\star\nu} = \delta_{\mu\nu},
\quad
\left[\pd \tilde{\vec{e}}_{\sigma}(\tilde{x})/\pd \tilde{\varrho}^\sigmap\right]_\star = 0.
\end{gather}
(Here and further, the indices $\sigma$ and $\sigmap$ span from 1 to 2; other Greek indices span from 1 to 3. The specific choice of $\tilde{\vec{e}}_\mu$ used in \parade is discussed in \Sec{sec:code}.) Then,
\begin{gather}
\vec{x} \approx \vec{X}(\zeta) + \left(
\begin{array}{cc}
\tilde{\vec{e}}_{\star 1} & \tilde{\vec{e}}_{\star 2}
\end{array}
\right)
\left(
\begin{array}{c}
\tilde{\varrho}^1\\
\tilde{\varrho}^2
\end{array}
\right)
\end{gather}
and
\begin{gather}\label{eq:mcXX}
\mc{X}^\alpha{}_\zeta \approx (\tilde{\vec{e}}_{\star\zeta})^\alpha + [(\tilde{\vec{e}}_{\star\sigma})^\alpha]'\tilde{\varrho}^\sigma,
\quad
\mc{X}^\alpha{}_\sigma \approx (\tilde{\vec{e}}_{\star\sigma})^\alpha,
\end{gather}
where the prime denotes the derivative with respect to $\zeta$. This leads to
\begin{gather}
\vec{\mc{X}} = \vec{\mc{X}}_\star + \Delta \vec{\mc{X}},
\quad
\Delta \vec{\mc{X}} \approx \vec{\mc{Y}}_{\star\sigma} \tilde{\varrho}^\sigma,
\end{gather}
where $\vec{\mc{Y}}_{\star\sigma}$ is a $3 \times 3$ matrix given by
\begin{gather}\label{eq:Y}
\vec{\mc{Y}}_{\star\sigma} = \left(
\begin{array}{ccc}
[(\tilde{\vec{e}}_{\star\sigma})^1]' & 0 & 0 \\[3pt]
[(\tilde{\vec{e}}_{\star\sigma})^2]' & 0 & 0 \\[3pt]
[(\tilde{\vec{e}}_{\star\sigma})^3]' & 0 & 0
\end{array}
\right).
\end{gather}
Using that $\vec{\mc{X}} \tilde{\vec{\mc{X}}} = \mathbb{1}$, we also obtain $\tilde{\vec{\mc{X}}} = \tilde{\vec{\mc{X}}}_\star + \Delta \tilde{\vec{\mc{X}}}$,
\begin{gather}
\Delta \tilde{\vec{\mc{X}}} \approx
- \tilde{\vec{\mc{X}}}_\star (\Delta \vec{\mc{X}}) \tilde{\vec{\mc{X}}}_\star
\approx
- \tilde{\vec{\mc{X}}}_\star \vec{\mc{Y}}_{\star\sigma}  \tilde{\vec{\mc{X}}}_\star \tilde{\varrho}^\sigma.\label{eq:dXY}
\end{gather}

For future references, note that we shall use tilde also to mark the components of vectors and tensors measured in these new coordinates. As usual, these components are connected to those in the laboratory coordinates via~\cite{book:landau2}
\begin{eqnarray*}
A^\alpha = \mc{X}^\alpha{}_\mu \tilde{A}^\mu,
& \quad &
A_\alpha = \tilde{A}_\mu \tilde{\mc{X}}^\mu{}_\alpha,\\
A^{\alpha\beta} = \mc{X}^\alpha{}_\mu\mc{X}^\beta{}_\nu \tilde{A}^{\mu\nu},
& \quad &
A_{\alpha\beta} = \tilde{A}_{\mu\nu} \tilde{\mc{X}}^\mu{}_\alpha\tilde{\mc{X}}^\nu{}_\beta,\\
A^\alpha{}_\beta = \mc{X}^\alpha{}_\mu \tilde{\mc{X}}^\nu{}_\beta \tilde{A}^\mu{}_\nu,
& \quad &
A_\alpha{}^\beta = \tilde{\mc{X}}^\mu{}_\alpha \mc{X}^\beta{}_\nu \tilde{A}_\mu{}^\nu.
\end{eqnarray*}
We also introduce first-order partial derivatives
\begin{gather}
f_{|\alpha} \doteq \frac{\pd f(\vec{x}, \vec{k})}{\pd x^\alpha},
\quad
f^{|\alpha} \doteq \frac{\pd f(\vec{x}, \vec{k})}{\pd k_\alpha},
\end{gather}
and the second-order derivatives are denoted as follows:
\begin{gather}\notag
f_{|\alpha\beta} \doteq \frac{\pd^2 f(\vec{x}, \vec{k})}{\pd x^\alpha\pd x^\beta},
\kern 5pt
f^{|\alpha\beta} \doteq \frac{\pd^2 f(\vec{x}, \vec{k})}{\pd k_\alpha\pd k_\beta},
\kern 5pt
f^{|\alpha}_{|\beta} \doteq \frac{\pd^2 f(\vec{x}, \vec{k})}{\pd k_\alpha\pd x^\beta}.
\end{gather}
Accordingly, $\tilde{f}_{|\mu} \doteq \pd f(\vec{x}, \vec{k})/\pd \tilde{x}^\mu$, and so on.

\subsection{Quasioptical equation}
\label{sec:qo}

\subsubsection{Basic formulas}

Using the fact that the metric in the assumed ray-based coordinates is Euclidean on the RR and overall smooth, the envelope equation can be represented as follows \cite{foot:paper1}:
\begin{multline}\label{eq:aeq1}
\Lambda a + i \Gamma a - i\tilde{V}^\mu \pd_\mu a - \frac{i}{2}\,(\pd_\mu \tilde{V}^\mu) a - U a
\\ - \frac{1}{2}\,\pd_\sigma(
\tilde{\Phi}^{\sigma\sigmap} \pd_{\sigmap}a)
= 0.
\end{multline}
Here, we introduced
\begin{gather}
\Gamma \doteq \vec{\Xi}^+ \vec{D}_A \vec{\Xi},\label{eq:Gamma}\\
\tilde{V}^{\mu} \doteq \tilde{\Lambda}^{|\mu},
\quad
\tilde{\Phi}^{\sigma\sigmap} \doteq \tilde{\Lambda}^{|\sigma\sigmap},\\
U \doteq \text{Im}\,(
\Lambda_{|\alpha} \vec{\Xi}^+ \vec{\Xi}^{|\alpha}
-\Lambda^{|\alpha} \vec{\Xi}^+ \vec{\Xi}_{|\alpha}
+ \vec{\Xi}^{+|\alpha} \vec{D}_H \vec{\Xi}_{|\alpha}
),\label{eq:U}
\end{gather}
and also the following notation for the derivatives:
\begin{gather}
\pd_\mu \equiv \frac{\pd}{\pd \tilde{x}^\mu}, \quad
\pd_\sigma \equiv \frac{\pd}{\pd \varrho^\sigma}, \quad
\pd_{\sigmap} \equiv \frac{\pd}{\pd \varrho^{\sigmap}}. \label{eq:grad}
\end{gather}
Note that all coefficients in \Eq{eq:aeq1} are considered as functions of $\vec{x}$\shout{;} namely, $f \equiv f(\vec{x}, \vec{k}(\vec{x}))$ for any $f$. Accordingly, the derivatives \eq{eq:grad} act on both arguments. Also note that $\Gamma$ and $U$ are calculated in the laboratory coordinates; otherwise, additional terms would emerge~\cite{foot:paper1}.

Within the assumed accuracy, $\tilde{\Phi}^{\sigma\sigmap}$ and $U$ can be replaced with those evaluated on the RR,
\begin{gather}
\tilde{\Phi}^{\sigma\sigmap}_\star =
\tilde{\mc{X}}_\star^\sigma{}_\alpha\tilde{\mc{X}}_\star ^{\sigmap}{}_\beta\Lambda_\star^{|\alpha\beta},
\label{eq:Phistar}
\\
U_\star = \text{Im}\,(
\Lambda_{|\alpha} \vec{\Xi}^+ \vec{\Xi}^{|\alpha}
-\Lambda^{|\alpha} \vec{\Xi}^+ \vec{\Xi}_{|\alpha}
+ \vec{\Xi}^{+|\alpha} \vec{D}_H \vec{\Xi}_{|\alpha}
)_\star.\label{eq:Ustar}
\end{gather}
In contrast, $\Lambda$, $\Gamma$, and $\tilde{V}^\mu$ generally must be calculated with a higher accuracy. (An example of a numerical simulation that ignores this fact is presented in \Sec{sec:bend} for a reference.) In order to do that, we need to find $\vec{k}(\vec{x})$, so we start by specifying $\theta$, through which $\vec{k}$ is defined (\Sec{sec:basic}). Let us adopt
\begin{gather}\label{eq:theta}
\theta(\zeta, \vec{\varrho}) = \int \tilde{K}_\zeta(\zeta)\,\dd\zeta + \tilde{K}_\sigma(\zeta) \tilde{\varrho}^\sigma.
\end{gather}
(This choice of $\theta$ is different from the one assumed in Paper~I but similar to that in \Ref{ref:balakin07a}.) Hence,
\begin{gather}
\tilde{k}_\sigma(\vec{x}) = \pd_\sigma \theta = \tilde{K}_\sigma(\zeta),\label{eq:aux1}\\
\tilde{k}_\zeta(\vec{x}) = \pd_\zeta \theta = \tilde{K}_\zeta(\zeta) + \tilde{K}'_\sigma(\zeta) \tilde{\varrho}^\sigma.\label{eq:aux2}
\end{gather}
To simplify this, note that
\begin{gather}
\dd\theta_\star = \tilde{K}_\mu \dd\tilde{X}^\mu = K_\alpha\dd X^\alpha
\end{gather}
by definition. Hence, the transformation $(X^\alpha, K^\alpha) \mapsto (\tilde{X}^\mu, \tilde{K}^\mu)$ is symplectic (canonical), so the ``tilded'' coordinates satisfy Hamilton's equations similar to \Eq{eq:rtpX},
\begin{gather}\label{eq:rtpt}
\frac{\dd \tilde{X}^\mu}{\dd\zeta} = \frac{\tilde{V}^\mu_\star}{V_\star},
\quad
\frac{\dd \tilde{K}_\mu}{\dd\zeta} = - \frac{\tilde{\Lambda}_{\star|\mu}}{V_\star}.
\end{gather}
By combining \Eqs{eq:aux1}, \eq{eq:aux2}, and \eq{eq:rtpt}, one arrives at
\begin{gather}
\tilde{k}_\mu(\vec{x}) = \tilde{K}_\mu(\zeta) + \tilde{\dk}_\mu,
\quad
\tilde{\dk}_\mu = - \tilde{\varrho}^\sigma \tilde{\Lambda}_{\star|\sigma} \tilde{V}_{\star\mu}/V_\star^2,
\label{eq:aux4}
\end{gather}
where we used $\tilde{V}_{\star\mu} = V_\star \delta_\mu^\zeta$, and $\delta_\mu^\zeta$ is a Kronecker symbol.

\subsubsection{Calculating $\Lambda$ and $\Gamma$}

Expanding $\Lambda(\vec{x}, \vec{k}(\vec{x}))$ around the RR, one obtains
\begin{multline}
\Lambda(\vec{x}, \vec{k}(\vec{x}))
= \Lambda_\star
+ \tilde{\Lambda}_{\star|\sigma} \tilde{\varrho}^\sigma + \tilde{\Lambda}^{|\mu}_\star \tilde{\dk}_\mu
+\frac{1}{2}\,\tilde{\Lambda}_{\star|\sigma\sigmap}\,\tilde{\varrho}^\sigma \tilde{\varrho}^{\sigmap}
\\
+ \tilde{\Lambda}_{\star|\sigma}^{|\mu}\,\tilde{\varrho}^{\sigma} \tilde{\dk}_{\mu}
+ \frac{1}{2}\,\tilde{\Lambda}^{|\mu\nu}_\star\,\tilde{\dk}_{\mu}\tilde{\dk}_{\nu}
+ ...
\end{multline}
Using \Eqs{eq:Lz} and \eq{eq:aux4}, one can reduce this down to
\begin{gather}
\Lambda(\vec{x}, \vec{k}(\vec{x}))
 \approx \frac{1}{2}\,\bigg(
\tilde{\Lambda}_{\star|\sigma\sigmap}\,\tilde{\varrho}^\sigma \tilde{\varrho}^{\sigmap}
+ 2\,\tilde{\Lambda}_{\star|\sigma}^{|\mu}\,\tilde{\varrho}^{\sigma} \tilde{\dk}_{\mu}
+ \tilde{\Lambda}^{|\mu\nu}_\star\,\tilde{\dk}_{\mu}\tilde{\dk}_{\nu}
\bigg),\notag
\end{gather}
which is of the order of $O(\epsilon_\perp^2) \sim O(\epsilon_\parallel)$, as needed.
Since the metric on the RR is Euclidean, the second-order derivatives entering this equation are transformed as true tensors. Hence, they can be mapped to the laboratory coordinates with $\vec{\mc{X}}_\star$ and $\tilde{\vec{\mc{X}}}_\star$. This leads to $\Lambda(\vec{x}, \vec{k}(\vec{x})) \approx \tilde{\mcc{L}}_{\star\sigma\sigmap} \tilde{\varrho}^\sigma\tilde{\varrho}^{\sigmap}$ with
\begin{multline}\label{eq:Thete}
\tilde{\mcc{L}}_{\star\sigma\sigmap} =
\frac{1}{2}
\bigg(
\Lambda_{\star|\alpha\beta}
- \frac{2V_{\star\gamma}}{V^2_\star}
\,\Lambda_{\star|\alpha}^{|\gamma}\Lambda_{\star|\beta}
\\
+ \frac{V_{\star\gamma}V_{\star\delta}}{V^4_\star}
\,\Lambda^{|\gamma\delta}_\star\Lambda_{\star|\alpha}\Lambda_{\star|\beta}
\bigg)
\mc{X}_\star^\alpha{}_{\sigma} \mc{X}_\star^\beta{}_{\sigmap},
\end{multline}
where we substituted [cf. \Eq{eq:aux4}]
\begin{gather}\label{eq:aux6}
\dk_\beta \approx - \varrho^\alpha \Lambda_{\star|\alpha} V_{\star\beta}/V_\star^2,
\quad
\varrho^\alpha \doteq \mc{X}_\star^\alpha{}_{\sigma} \tilde{\varrho}^\sigma.
\end{gather}
(Note that, to the leading order, $\vec{\varrho} \equiv \{\varrho^1, \varrho^2\}$ can be considered as a true vector and transforms accordingly.)

Using \Eqs{eq:aux6}, one can also adopt the following approximation of \Eq{eq:Gamma} for $\Gamma$:
\begin{gather}\label{eq:Gammastar}
\Gamma(\vec{x}, \vec{k}(\vec{x})) = \vec{\Xi}^+_\star \vec{D}_A(\vec{X}(\zeta) + \vec{\varrho}, \vec{K}(\zeta) + \vec{\dk}(\zeta, \vec{\varrho})) \vec{\Xi}_\star.
\end{gather}
Here, the polarization vectors $\vec{\Xi}$ and $\vec{\Xi}^+$ are evaluated at the RR (hence the stars), because they are determined by $\vec{D}_H$, which is a slow function. In contrast, $\vec{D}_A$ can vary substantially \shout{within} the beam cross-section, so it can be important to calculate $\vec{D}_A$ separately at each point.

\subsubsection{Calculating $\tilde{V}$}

In order to calculate the terms involving $\tilde{V}^\mu = \tilde{\mc{X}}^\mu{}_\alpha V^\alpha$, let us represent this function as $\tilde{V}^\mu = \tilde{V}_\star^\mu + \tilde{v}^\mu$,
\begin{gather}
\tilde{v}^\mu(\vec{x}) =
(\Delta\tilde{\mc{X}}^{\mu}{}_\alpha)V^\alpha_\star
+ \tilde{\mc{X}}_\star^{\mu}{}_\alpha (\Delta V^\alpha),
\end{gather}
where $\Delta$ denotes deviations from the corresponding values of the RR. Using \Eq{eq:dXY}, we obtain
\begin{gather}
(\Delta\tilde{\mc{X}}^{\mu}{}_\alpha)V^\alpha_\star = - (\tilde{\vec{\mc{X}}}_\star \vec{\mc{Y}}_{\star\sigma}  \tilde{\vec{\mc{X}}}_\star \vec{V}_\star)^\mu \tilde{\varrho}^\sigma.
\end{gather}
Also note that
\begin{align}
\Delta V^\alpha
& \approx \Lambda^{|\alpha}(\vec{x}, \vec{k}(\vec{x})) - \Lambda^{|\alpha}_\star
\notag \\
& \approx \Lambda^{|\alpha}_{\star|\beta} \varrho^\beta + \Lambda^{|\alpha\gamma}_\star\dk_\gamma
\notag \\
& = [\Lambda^{|\alpha}_{\star|\beta} - (V_{\star\gamma}/V^2_\star)\,\Lambda^{|\alpha\gamma}_\star
\Lambda_{\star|\beta}]\varrho^\beta\shout{.}
\end{align}
Hence, $\tilde{v}^\mu(\vec{x}) \approx \tilde{\vartheta}_\star^\mu{}_{\sigmap} \tilde{\varrho}^{\sigmap}$, where
\begin{multline}
\label{eq:chi}
\tilde{\vartheta}_\star^\mu{}_{\sigmap} = \left[\Lambda^{|\alpha}_{\star|\beta}
- (V_{\star\gamma}/V^2_\star)\,\Lambda^{|\alpha\gamma}_\star
\Lambda_{\star|\beta}\right]\tilde{\mc{X}}^{\mu}_\star{}_\alpha \mc{X}^\beta_\star{}_{\sigmap}
\\- (\tilde{\vec{\mc{X}}}_\star \vec{\mc{Y}}_{\star\sigmap}  \tilde{\vec{\mc{X}}}_\star \vec{V}_\star)^\mu.
\end{multline}
Then, since $\tilde{V}^\sigma_\star = 0$ and $\tilde{v}^\zeta \pd_\zeta = O(\epsilon_\parallel)$, one obtains
\begin{gather}
\tilde{V}^\mu \pd_\mu a \approx V_\star \,\pd_\zeta a + \tilde{v}^\sigma \pd_\sigma a,
\quad
\pd_\mu\tilde{V}^\mu \approx V'_\star(\zeta) + \tilde{\vartheta}_\star^\sigma{}_\sigma,
\end{gather}
where $\tilde{\vartheta}_\star^\sigma{}_\sigma$ can be calculated from \Eq{eq:chi} by taking $\mu = \bar{\sigma} = \sigma$ and summing over $\sigma$ accordingly. Also,
\begin{gather}\label{eq:vstar}
V_\star = |\pd_{\vec{K}} \Lambda(\vec{X}, \vec{K})|.
\end{gather}

\subsubsection{Final result}
\label{sec:final}

Using the above results, one can express \Eq{eq:aeq1} as
\begin{multline}\label{eq:ee}
\tilde{\mcc{L}}_{\star\sigma\sigmap} \tilde{\varrho}^\sigma\tilde{\varrho}^{\sigmap} a
+ i \Gamma a
- iV_\star \pd_\zeta a
- i \tilde{\vartheta}_\star^\sigma{}_{\sigmap} \tilde{\varrho}^{\sigmap} \pd_\sigma a
\\
- \frac{i}{2}\,(V'_\star + \tilde{\vartheta}_\star^\sigma{}_\sigma) a
- U_\star a
- \frac{1}{2}\,\tilde{\Phi}^{\sigma\sigmap}_\star \pd^2_{\sigma\sigmap} a
= 0,
\end{multline}
which also has the following corollary:
\begin{gather}\label{eq:cons}
\frac{\dd P}{\dd\zeta}  = 2 \int \Gamma \,|a|^2 \,\dd^2\tilde{\varrho},\quad P \doteq V_\star \int |a|^2 \,\dd^2\tilde{\varrho}.
\end{gather}
Equation \eq{eq:cons} shows that $P$ is conserved when $\Gamma$ is zero. As explained in Paper~I, this represents the conservation of the wave action, and $P$ equals the energy flux up to a constant coefficient. Assuming the notation
\begin{gather}\label{eq:aphi}
\phi \doteq \sqrt{V_\star} a,
\end{gather}
so $P = \int |\phi|^2 \,\dd^2\tilde{\varrho}$, one can also rewrite \Eq{eq:ee} as
\begin{multline}\label{eq:FE}
V_\star \pd_\zeta \phi =
- \tilde{\vartheta}_\star^\sigma{}_{\sigmap} \tilde{\varrho}^{\sigmap} \pd_\sigma \phi
- \frac{\tilde{\vartheta}_\star^\sigma{}_\sigma}{2}\,\phi
+ \Gamma \phi
\\
- i(\tilde{\mcc{L}}_{\star\sigma\sigmap} \tilde{\varrho}^\sigma\tilde{\varrho}^{\sigmap} - U_\star) \phi
+  \frac{i}{2}\,\tilde{\Phi}^{\sigma\sigmap}_\star\pd^2_{\sigma\sigmap} \phi.
\end{multline}

Equation \eq{eq:FE} is the main quasioptical equation used in \parade. As a reminder, we assume \Eq{eq:vstar} for $V_\star$, \Eq{eq:Gammastar} for $\Gamma$, \Eq{eq:Thete} for $\tilde{\mcc{L}}_{\star\sigma\sigmap}$, \Eq{eq:Ustar} for $U_\star$, and \Eq{eq:Phistar} for $\tilde{\Phi}^{\sigma\sigmap}_\star$. We also assume \Eq{eq:chi} for $\tilde{\vartheta}_\star^\sigma{}_{\sigmap}$, with \Eq{eq:Y} for $\vec{\mc{Y}}_\star$. The quantity $\tilde{\vartheta}_\star^\sigma{}_\sigma$ can be calculated from \Eq{eq:chi} by taking $\mu = \bar{\sigma} = \sigma$ and summing over $\sigma$ accordingly. The matrix $\vec{\mc{X}}$ is given by \Eq{eq:mcXX}, and the matrix $\tilde{\vec{\mc{X}}}$ is given by \Eq{eq:tX}. The index $\star$ denotes that the corresponding quantities are evaluated at the RR (\ie at $\tilde{\vec{\varrho}} = 0$) and thus are independent of the transverse coordinates.

\section{Numerical algorithm}
\label{sec:code}

\begin{figure*}
\begin{center}
\includegraphics[width=14.0cm,clip]{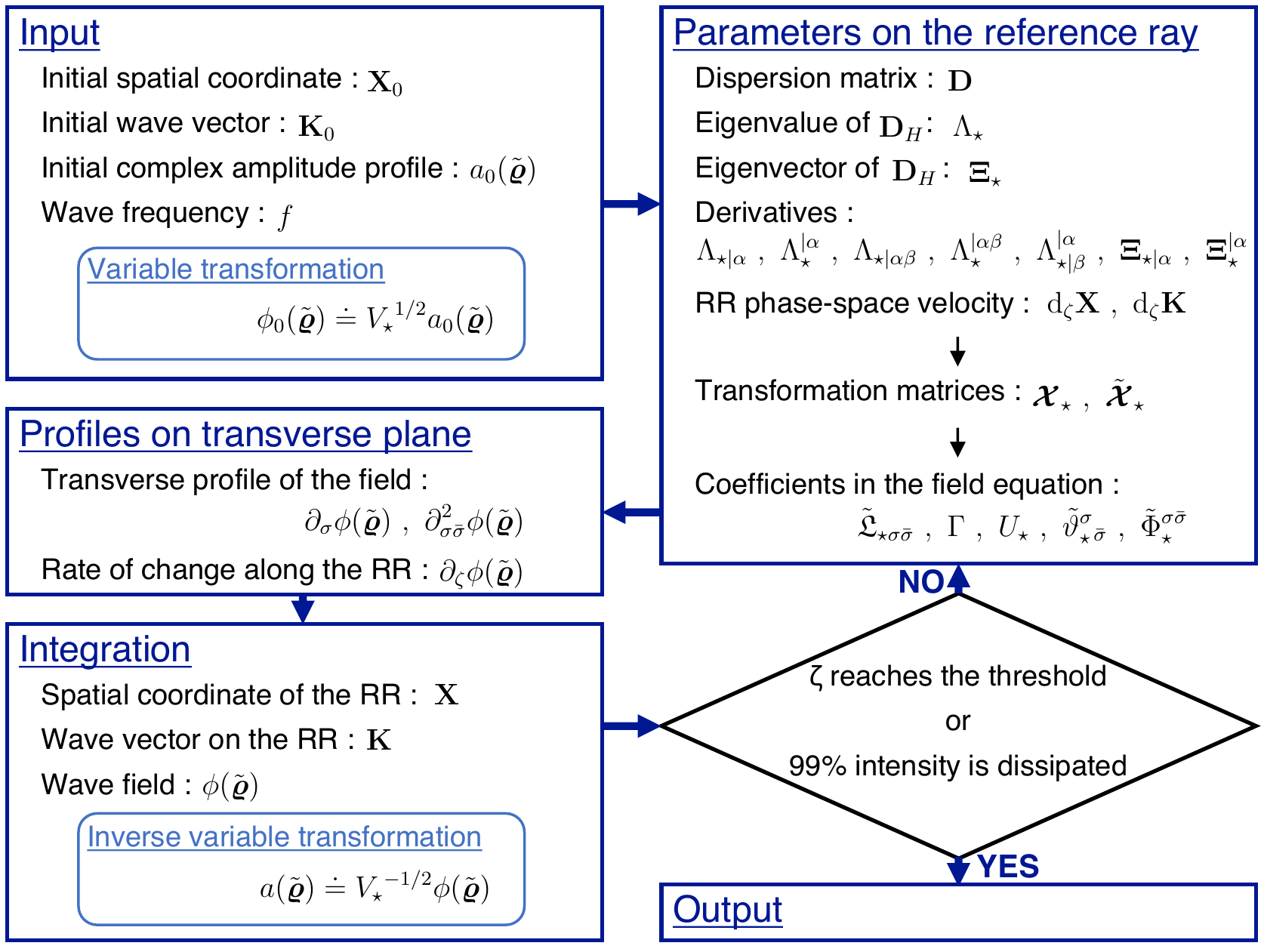}
\end{center}
\caption{A flowchart outlining the operation of \parade.}
\label{fig:scheme}
\end{figure*}

The code \parade is based on the model described in \Sec{sec:xgo}. The RR starting point $\vec{X}_0$, the orientation of the RR initial wave vector, the frequency $f \doteq \omega/2\pi$, the initial profile of the complex amplitude $a_0$, the magnetic field profile $\vec{B}(\vec{x})$, and the plasma density profile $n(\vec{x})$ are considered pre-determined. Currently, they are entered as analytic functions, but it is also possible to use experimental data or profiles generated numerically. The value of $|\vec{K}_0|$ is determined by the local dispersion relation at $\vec{X}_0$. The collisionless cold-electron-plasma model is assumed for the dielectric tensor \cite{book:stix} here, but the code can also accept a general dispersion tensor $\vec{D}$. In particular, warm-plasma simulations will be reported separately.

Figure~\ref{fig:scheme} outlines the code operation. At each point on the RR, starting from the initial point $\vec{X}_0$, the dispersion tensor $\vec{D} = \vec{D}_H + i\vec{D}_A$ is calculated. Then, the eigenvalues of $\vec{D}_H$ are calculated, and the eigenvalue $\Lambda_\star$ is found that has the least absolute value. This $\Lambda_\star$ is then numerically differentiated and used as the Hamiltonian in \Eqs{eq:rtpX} to propagate the RR using the fourth-order Runge--Kutta method. Based on the knowledge of the RR trajectory $\vec{X}(\zeta)$, the ray-based coordinates $\tilde{x}^\mu$ are constructed by choosing the basis vectors $\tilde{\vec{e}}_{\star\mu}$, which is done as follows.

We adopt $\tilde{\vec{e}}_{\star\zeta}$ to be the unit vector along the RR group velocity. The other two vectors $\tilde{\vec{e}}_{\star\sigma}$ (where $\sigma = 1, 2$) can, in principle, be chosen arbitrarily as long as they satisfy \Eqs{eq:eee} and the resulting metric is smooth. The traditional Frenet--Serret frame can be used but sometimes fails the latter requirement. Instead, we choose
\begin{gather}
\tilde{\vec{e}}_{\star\zeta} = \frac{\vec{X}'}{|\vec{X}'|},
\\
\tilde{\vec{e}}_{\star 1} = \frac{\vec{X}' \times (\tilde{\vec{e}}_{\star 1}^\bullet \times \vec{X}')}{|\vec{X}' \times (\tilde{\vec{e}}_{\star 1}^\bullet \times \vec{X}')|},
\\
\tilde{\vec{e}}_{\star 2} = \frac{\vec{X}' \times \tilde{\vec{e}}_{\star 1}^\bullet}{|\vec{X}' \times \tilde{\vec{e}}_{\star 1}^\bullet|}
\end{gather}
at each given time step, where the vectors marked with~$^\bullet$ are those from the previous time step. (The prime denotes the derivative with respect to $\zeta$, so $|\vec{X}'| = 1$.) For the very first time step, $\tilde{\vec{e}}_{\star 1}^\bullet$ and $\tilde{\vec{e}}_{\star 2}^\bullet$ can be chosen arbitrarily as long as they are orthogonal to each other and to the initial group velocity on the RR. This ensures that the basis vectors $\tilde{\vec{e}}_{\star\mu}$ transform continuously along the RR and result in a smooth ray-based metric.

As the next step, the polarization eigenvector $\vec{\Xi}_\star$ is calculated as the eigenvector of $\vec{D}_H$ corresponding to the eigenvalue $\Lambda_\star$; also, its complex conjugate $\vec{\Xi}^+_\star$ is determined. After these vectors are numerically differentiated, the coefficients in \Eq{eq:FE} are calculated as outlined in \Sec{sec:final}. The field $a$ is mapped to $\phi$ using \Eq{eq:aphi}, and $\phi$ is differentiated with respect to $\tilde{\varrho}^\sigma$ at fixed $\zeta$ using the finite-difference approach. Then, $\phi$ is propagated by solving \Eq{eq:FE} using the fourth-order Runge--Kutta method. Finally, $\phi$ is mapped back to $a$, and the three-dimensional profile $a(\vec{x})$ is yielded as a data array.

\section{Simulation results}
\label{sec:sim}

In this section, test simulation results are reported. These results were obtained by applying \parade to sample problems that fit the assumptions of the model described above. \shout{(}Simulations that involve mode conversion are presented in Paper~III.\shout{) All simulations were done on a laptop with Intel Core\textsuperscript{TM}~i7-7660U processor and took only a few seconds to run, as further specified in the figure captions.}
\subsection{Beams in vacuum}
\label{sec:vacuum}

\begin{figure*}
\begin{center}
\includegraphics[width=15.0cm,clip]{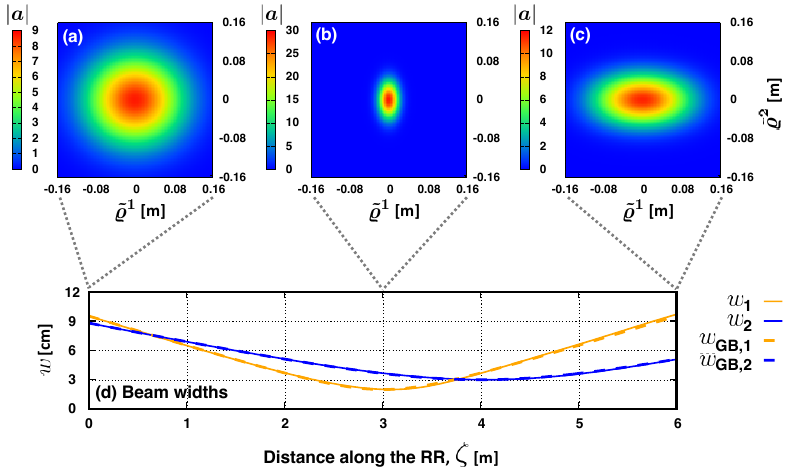}
\end{center}
\caption{\parade simulation of a wave beam in vacuum. Shown is the transverse intensity profile at different locations along the beam trajectory: (a) $\zeta = 0.0$~m, (b) $\zeta = 3.0$~m, and (c) $\zeta = 6.0$~m. (d) shows the corresponding beam widths $w_1$ and $w_2$ along the two orthogonal transverse axes. \shout{The computing time is approximately $4.4$~s.} The analytic solution for $w_{1,2}(\zeta)$ from Gaussian-beam optics is also presented for comparison (denoted $w_{\rm GB, 1}$ and $w_{\rm GB, 2}$, respectively).
}
\label{fig:vacuum}
\end{figure*}

As the first example, we simulated the propagation of a wave beam in vacuum. For the initial beam profile, we assumed a Gaussian profile
\begin{multline}\label{eq:gb}
a_0 = \sqrt{\frac{2}{\pi w_1 w_2}}\,\exp
\bigg[
- \frac{(\tilde{\varrho}^1)^2}{{w_1}^2} - \frac{(\tilde{\varrho}^2)^2}{{w_2}^2}
\\
+ \frac{i k (\tilde{\varrho}^1)^2}{2 R_1} + \frac{i k (\tilde{\varrho}^2)^2}{2 R_2}
+ \frac{i}{2} (g_1 + g_2)
\bigg],
\end{multline}
where
\begin{gather}
w_\mu \doteq w_{0,\mu} \sqrt{1 + \varsigma^{-2}},
\quad
R_\mu \doteq \mc{Z}_\mu (1 + \varsigma^2),
\\
g_\mu \doteq \tan^{-1} \varsigma,
\quad
\varsigma \doteq k w_{0,\mu}/(2\mc{Z}_\mu),
\end{gather}
and $k = 2\pi f/c$, with the focal lengths $\mc{Z}_1 = 3.0$~m and $\mc{Z}_2 = 4.0$~m, the waist sizes $w_{0,1} = 2.0$~cm and $w_{0,2} = 3.0$~cm, and the wave frequency $f = 154.0$~GHz, which corresponds to the vacuum wavelength $\lambda_0 \approx 2$~mm. The results are shown in \Fig{fig:vacuum}. At $\zeta > 0$, the beam width $w_\sigma$ is determined numerically as the distance on which $|a|$ drops one $e$-fold from its maximum along the $\tilde{\varrho}^\sigma$-axis. As seen from \Fig{fig:vacuum}(d), \parade demonstrates good agreement with the analytical formula from Gaussian-beam optics \cite{book:yariv}.

\subsection{Straight beams in cold plasma}
\label{sec:straight}

\begin{figure}
\begin{center}
\includegraphics[width=8.2cm,clip]{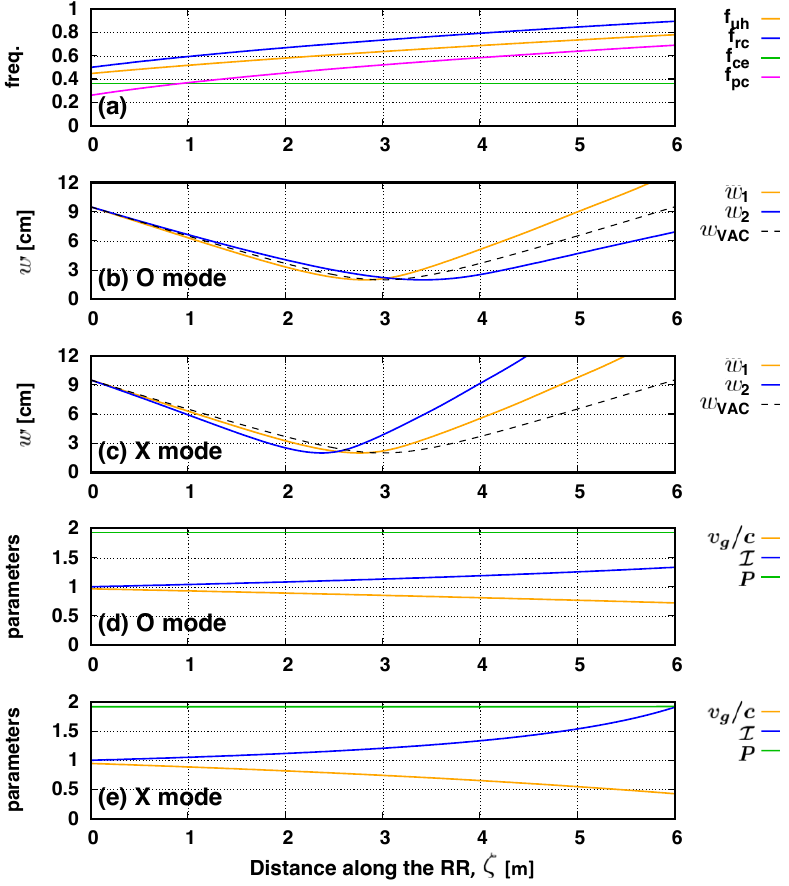}
\end{center}
\caption{\parade simulation of a straight wave beam propagating perpendicular to a homogeneous magnetic field and parallel to the density gradient [\Eq{eq:nB}] in cold electron plasma. (a) shows how the key frequencies that determine the wave dispersion evolve along the RR; specifically, shown are the upper-hybrid frequency $f_{\rm uh}$, the right-cutoff frequency $f_{\rm rc}$, the electron cyclotron frequency  $f_{\rm ce}$, and the plasma frequency $f_{\rm pc}$, all in units $f$. (b) and (c) show the beam widths for the O~mode and the X~mode, respectively. To demonstrate the importance of plasma dispersion, simulation results for a wave beam in vacuum are also presented as a reference. The notation is the same as in \Fig{fig:vacuum}. (d) and (e) show the group velocity \shout{$v_g \doteq V_\star / |\pd_{\omega} \Lambda_\star|$} in units $c$, \shout{the integrated normalized intensity $\mathcal{I} \doteq \int |a|^2 \,\dd^2\tilde{\varrho}$}, and the wave-energy flux \shout{$P = V_\star \mathcal{I}$} [see \Eq{eq:cons}] in arbitrary units for the O~mode and the X~mode, respectively. \shout{The computing time is approximately $12$~s.}}
\label{fig:slab}
\end{figure}

As the second example, we simulated the propagation of waves in collisionless magnetized cold electron plasma. From now on, we introduce the standard notation $\{x, y, z\}$ for the laboratory coordinates and $\vec{e}_\alpha$ for the unit vectors along the corresponding axes. In particular, the $x$ axis is chosen to be along the initial direction of the wave vector. We assume a slab geometry with
\begin{gather}\label{eq:nB}
n = n_0 (1 + x/L_n), \quad \vec{B} = \vec{e}_z B_0.
\end{gather}
Here, $n_0 = 2.0 \times 10^{19}$~m$^{-3}$, $L_n = 1.0$~m, and $B_0 = 2.0$~T. The initial field has the Gaussian profile \eq{eq:gb} with the focal lengths $\mc{Z}_1 = \mc{Z}_2 = 3.0$~m, the waist sizes $w_{0,1} = w_{0,2} = 2.0$~cm, and the wave frequency $f = 154.0$~GHz.
The simulation results are presented in \Fig{fig:slab}. Since the waves propagate \textit{along} the gradient of $n$, the RR trajectory is straight and does not depend on the polarization, just like in the vacuum case. However, the intensity profiles of the waves with different polarizations, which are the O and X waves here, are different from each other and from those in vacuum. The X wave is affected by the plasma more significantly. This is explained by the fact that the wave frequency $f$ is closer to the X-wave cutoff frequency $f_{\rm rc}$ than to the O-wave cutoff frequency $f_{\rm pc}$.
In addition, although $\mc{Z}_1$ and $\mc{Z}_2$ are assumed to be equal initially [dashed line in \Fig{fig:slab}(b) and (c)], they start to differ within the plasma [solid lines in \Fig{fig:slab}(b) and (c)] due to the magnetic field. Also, while the focal lengths are different from those in vacuum, the waist sizes are, in fact, the same. This is because $\vec{k}$ is uniform in the transverse plane for the lack of the density gradient in that plane. Simulations with transverse gradients are presented in \Sec{sec:bend}.
Then, we have also tested \parade for the energy conservation [\Eq{eq:cons}], which is expected here because waves in collisionless cold plasma have $\Gamma = 0$. While the wave intensity and group velocity change considerably along the RR path, the wave-energy flux is conserved by the code with high accuracy, as seen in \Figs{fig:slab}(d) and (e).

\subsection{Bended beams in cold plasma}
\label{sec:bend}

\begin{figure}
\begin{center}
\includegraphics[width=8.2cm,clip]{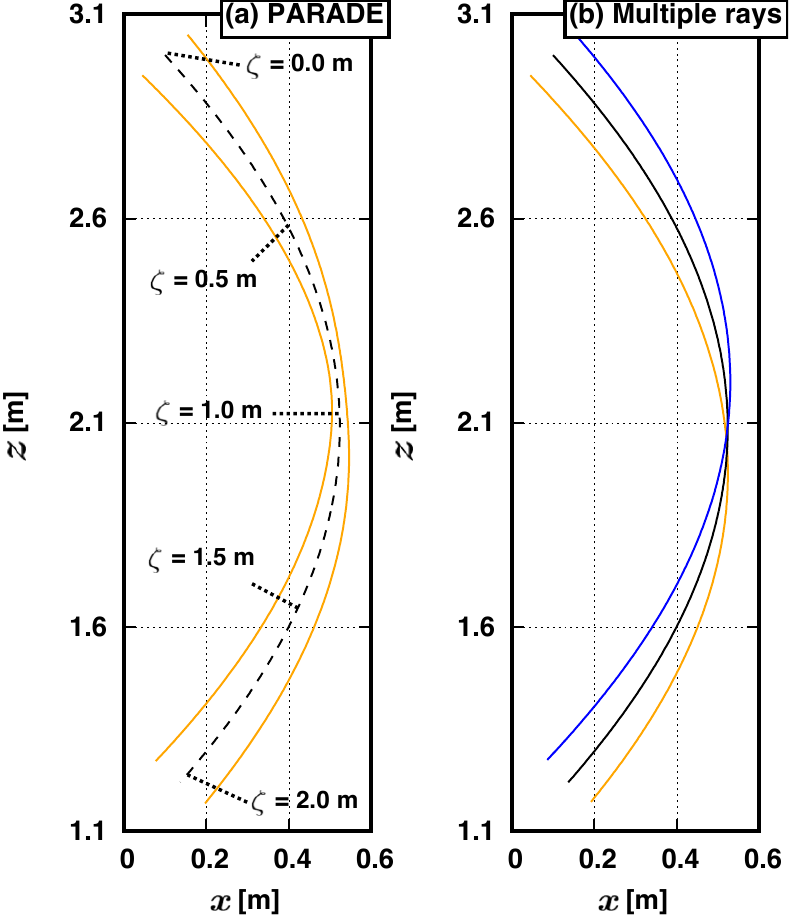}
\end{center}
\caption{(a) shows \parade simulation of a bended X wave beam in inhomogeneous plasma with inhomogeneous magnetic field [\Eqs{eq:nB21} and \eq{eq:nB22}]. \shout{The computing time is approximately $15$~s.} The dashed curves show the RR trajectory; the solid curves show the beam width as defined in \Sec{sec:vacuum}. \shout{Also indicated are the locations on the RR that correspond to the specific values of $\zeta$ used for the different subfigures in \Fig{fig:fit}.} (b) shows ray tracing simulations of the same beam as in (a). Unlike in the \parade simulation, here, diffraction is neglected and each ray propagates independently.}
\label{fig:bend}
\end{figure}

As another example, we applied \parade to simulate the propagation of bended wave beams launched at generic angles relative to the density gradient and the magnetic field. The assumed density and magnetic-field profiles are as follows:
\begin{gather}
n = n_0 \exp \left[ - \frac{(x - x_c)^2}{L_n^2}\right],\label{eq:nB21}
\\
\vec{B} = \vec{e}_z B_0 \exp \left[ - \frac{(x - x_c)^2}{L_b^2} \right],\label{eq:nB22}
\end{gather}
where $n_0 = 5.0 \times 10^{19}$~m$^{-3}$, $x_c = 2.0$~m, $L_n = 2.0$~m, $B_0 = 2.0$~T, and $L_b = 2.0$~m. The RR starting point is $\{x,y,z\}=\{0.1 ,0.0 ,3.0\}$~m, and the target point used to fix the orientation of the RR initial wave vector is $\{x,y,z\}=\{3.1 ,0.0 ,0.0\}$~m. For the initial profile, we adopted the Gaussian profile \eq{eq:gb} with the focal lengths $\mc{Z}_1 = \mc{Z}_2 = 1.5$~m, the waist sizes $w_{0,1} = w_{0,2} = 2.7$~cm, and the wave frequency $f = 77.0$~GHz (or $\lambda_0 \approx 4$~mm).
A simulation of the X-mode propagation is shown in \Fig{fig:bend}(a). Since \parade accounts for diffraction, the beam retains a nonzero width on the whole trajectory. This is different from conventional ray tracing, where diffraction is neglected and each ray propagates independently. A comparison is presented in \Fig{fig:bend}(b). The difference is essential for resonant-plasma heating and current drive, where the \shout{wave} energy deposition must be calculated accurately, as also discussed in \Refs{ref:mazzucato89, ref:nowak93, ref:peeters96, ref:farina07, ref:pereverzev98, ref:poli01b, ref:poli01, ref:poli18, ref:balakin08b, ref:balakin07a, ref:balakin07b}.

\begin{figure}
\begin{center}
\includegraphics[width=8.2cm,clip]{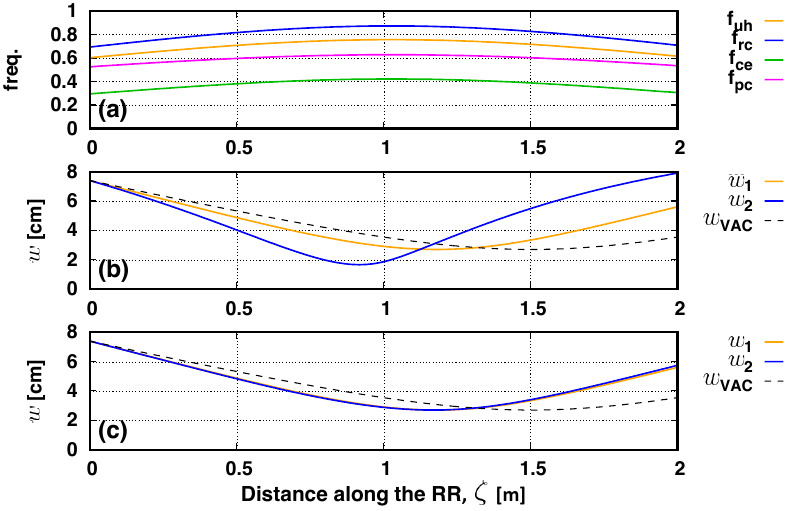}
\end{center}
\caption{\parade simulation of a bended X wave beam, same as in \Fig{fig:bend}. (a) shows the key frequencies along the RR trajectory. (b) shows the beam widths. For a reference, (c) illustrates the same simulation where the $\tilde{\vec{\varrho}}$-dependence of $\vec{k}$ is neglected. The dashed lines in (b) and (c) show the corresponding vacuum solutions. Note that both the focal lengths and waist sizes in (b) are considerably different from those in (c) and vacuum solutions.}
\label{fig:C}
\end{figure}

The corresponding evolution of the key frequencies along the RR is presented in \Fig{fig:C}(a), and the corresponding beam widths are presented in \Fig{fig:C}(b). For a reference, \Fig{fig:C}(c) shows the same widths calculated with neglected $\smash{\tilde{\vec{\varrho}}}$-dependence of $\vec{k}$. (In the latter case, the waist sizes are almost the same as for vacuum waves.) These results show that capturing the transverse inhomogeneity of $\vec{k}$ in practical simulations can be important.

\begin{figure}
\begin{center}
\includegraphics[width=8.2cm,clip]{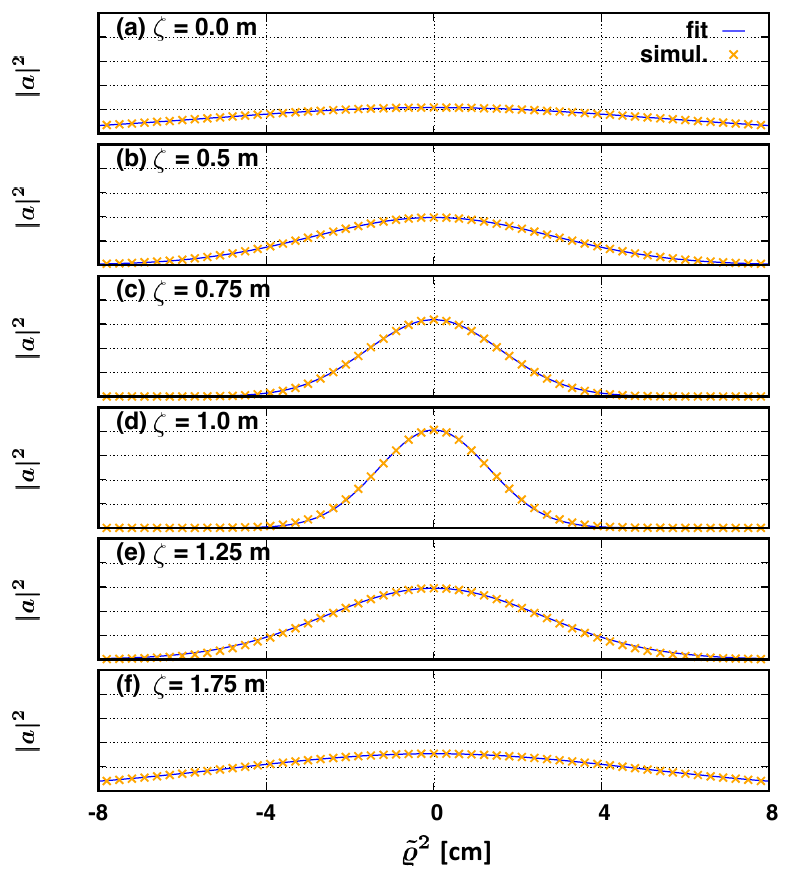}
\end{center}
\caption{The transverse intensity profile of a bended X wave beam, same as in \Fig{fig:bend}. The figures (a)-(f) correspond to $\zeta = 0.0$, $\zeta = 0.5$, $\zeta = 0.75$, $\zeta = 1.0$, $\zeta = 1.25$, and $\zeta = 1.75$~m, respectively. Shown in crosses are the results of \parade simulations of $|a|^2$ (in arbitrary units) versus $\tilde{\varrho}^2$ at fixed $\tilde{\varrho}^1 = 0$. The solid curves are Gaussian fits $|a|^2 = A \exp [-(\tilde{\varrho}^1/w_1)^2 - (\tilde{\varrho}^2/w_2)^2]$, where $w_1$, $w_2$, and $A$ were fitted numerically. Similar figures (not shown) have been also produced for $|a|^2$ versus $\tilde{\varrho}^1$ at fixed $\tilde{\varrho}^2 = 0$.
}
\label{fig:fit}
\end{figure}

We also note that in our cold-plasma \parade simulations, the beams remain Gaussian with high accuracy (\Fig{fig:fit}). However, it is generally not the case when resonant dissipation is included, for such dissipation can be strongly inhomogeneous in the beam cross-section. \parade simulations of dissipative beams in hot plasma will be reported in a separate publication.

\section{Conclusions}
\label{sec:conc}

We report the first quasioptical simulations of \shout{wave beams} in cold magnetized plasma using a new code \parade. This code does not assume any particular beam structure but instead solves a parabolic equation for the wave envelope; hence, it is particularly well-suited to modeling strongly inhomogeneous resonant dissipation in the future. The general theoretical model underlying \parade was largely derived in the previous paper \cite{foot:paper1} and describes multi-mode beams that can experience mode conversion. Here, we consider a simplified version of this theory in which mode conversion is not considered and a beam can be described as a scalar field. We adjust the theory to numerical modeling and simulate the propagation of wave beams in vacuum and inhomogeneous magnetized plasma. Our numerical results show good agreement with Gaussian-beam optics and surpass conventional ray tracing. Quasioptical simulations of mode-converting beams are reported in the next, third paper of this series.

\section{Acknowledgments}
The work was supported by the U.S. DOE through Contract No. DE-AC02-09CH11466. The work was also supported by JSPS KAKENHI Grant Number JP17H03514.


\begin{thebibliography}{10}
\bibitem{foot:balakincomp}
A comprehensive comparison of our model with those presented in
  \Refs{ref:balakin07a, ref:balakin07b, ref:balakin08b} would be cumbersome and
  is not attempted here also for the lack of detailed derivations in the
  mentioned papers. The verification of our model relies instead on the
  numerical results presented here and in Paper III, and also on the XGO
  applications to several test problems reported earlier \cite{phd:ruiz17,
  my:covar, my:xo, my:qdirac, my:qdiel}.
\end{thebibliography}

\end{document}